\documentclass{article}

\usepackage{arxiv}

\usepackage[utf8]{inputenc} 
\usepackage[T1]{fontenc}    
\usepackage{hyperref}       
\usepackage{url}            
\usepackage{booktabs}       
\usepackage{amsfonts}       
\usepackage{nicefrac}       
\usepackage{microtype}      
\usepackage{lipsum}
\usepackage{graphicx}
\usepackage{tabulary,graphicx,amsmath,amssymb,amsbsy,dcolumn,bm}
\usepackage{caption}
\usepackage{subcaption}
\DeclareMathOperator{\Tr}{Tr}
\usepackage[table,xcdraw]{xcolor}
\usepackage{url,multirow,morefloats,floatflt,cancel,tfrupee}
\usepackage{colortbl}
\usepackage{xcolor}
\usepackage{pifont}
\usepackage[nointegrals]{wasysym}
\title{Detecting communities via edge Random Walk Centrality}

\author{
 Ashwat Jain \\
  Wadham College, University of Oxford, United Kingdom \\
  \texttt{ashwat.jain@wadham.ox.ac.uk} \\
   \And
 P. Manimaran\\
  School of Physics, University of Hyderabad, India \\
  \texttt{manimaran@ouhyd.ac.in} \\
}

\begin{document}
\maketitle
\begin{abstract}
Herein we present a novel approach of identifying community structures in complex networks. We propose the usage of the Random Walk Centrality (RWC), first introduced by Noh and Rieger [Phys. Rev. Lett. 92.11 (2004): 118701]. We adapt this node centrality metric to an edge centrality metric by applying it to the line graph of a given network. A crucial feature of our algorithm is the needlessness of recalculating the centrality metric after each step, in contrast to most community detection algorithms. We test our algorithm on a wide variety of standard networks, and compare them with pre-existing algorithms. As a predictive application, we analyze the Indian Railway network for robustness and connectedness, and propose edges which would make the system even sturdier.
\end{abstract}


\section{Introduction}
Networks are ubiquitous - any interaction between members of a set may be represented so. The set may consist of humans (interactions then could take forms such as co-authoring a scientific paper \cite{coauthoringReview}, following on social networking sites \cite{socialMediaNetworks}, starring in the same movie \cite{coactorNetwork}, friendship \cite{friendshipNetwork}, internet connections \cite{Internet}, etc.), animals \cite{animalNetworks, DolphinNetwork} (in which case the interactions could be being part of the same pack/group, being preyed on/preying on each other, sharing a habitat with each other), inanimate objects like railway stations \cite{IndianRailwayAnalysis} (with two stations being called interacting if a train stops at them both), or any other entity. The members of the set are represented as 'nodes' or 'vertices' and the interactions between them as 'edges'. The resulting structure is what would be called a 'graph' or a 'network'.
\par

In addition to being extensively applicable to the fields of applied mathematics and statistical physics \cite{Strogatz, Barabasi, Dorogovtsev}, networks yield themselves as an excellent tool for analysis and modelling of interactions in the real world. They possess several interesting properties. One particular feature found in most networks is the presence of tightly-clustered subsets of nodes, called 'communities' \cite{FortunatoReview}. While there is no universally agreed-upon definition of a community \cite{Definitions}, it may be intuitively understood as a subset of the node set which consists of nodes which interact more amongst themselves than with those outside the community.
\par

Communities can tell us a lot about the network: if there exists a natural division amongst members, if members have a common choice, or even if they act collectively. Detection of community structure thus becomes an endeavor worth pursuing, and over the years we have seen many pioneering works \cite{GirvanNewman2002, Newman2004, Newman2004-2, Newman2006, Newman2008, Fortunato2007} on the subject. 
\par

Broad classification of community structure division algorithms is that into agglomerative and divisive clustering methods. In the former, we start with an unconnected set of nodes and progressively add links between them until a satisfactory community structure is reached. In the latter, we start with the original (connected) network and iteratively remove edges such that the remainder depicts communities. 
\par

It is evident that the usage of both agglomerative and divisive clustering methods requires the selection of an edge to add and remove respectively. The question then arises: given a network, how should one choose an edge for the application of clustering methods? The answer lies in centrality metrics - measures of how important a component of the graph is. The component may be an edge or a vertex, and in this case, we would like an edge centrality metric. We require a procedure to assign a value to each edge of a graph, and then we shall be able to choose the most/least important edge to remove/add during our clustering. 
\par

Numerous centrality metrics exist. A long (and possibly extensible) list may be found at \cite{MasterCentralities}. Several centrality metrics have previously been used to approach the problem of community structure detection. Consider, for example, Information Centrality \cite{InformationCentrality}, a metric based on the efficiency of information transfer over a network - and used for community structure identification in \cite{FortunatoLatoraMarchiori}. Another metric, the Resistance Distance, was used in \cite{ViaResistanceDistance} (originally introduced in \cite{ResistanceDistance}), which considered each edge as a fixed resistor and used the effective resistance between two nodes as a distance metric on the graph. In contrast, \cite{ViaCircuits} uses the same notion of electrical circuits, but defines distance based on the voltage difference between nodes. Several 'betweenness' metrics (namely Shortest Path, Random Walk and Current Flow) were introduced and used by \cite{GirvanNewman2004}. Another paper \cite{ViaRandomWalk1} uses a different random-walk based metric, called the diffusion distance: "The diffusion distance between two nodes is small if random walkers starting at these two nodes are likely to be at the same location at time t". Similarly, another notion of distance (i.e., a metric) is defined in \cite{ViaRandomWalk2} by claiming that random walks on a network get 'trapped' in communities. Many more methods, including several based on random walks may be found in the review article \cite{FortunatoReview}.
\par

A useful tool to interrelate node and edge centralities is that of a line graph. Given any network, an alternate and equivalent network can be constructed with nodes of the new network representing edges of the old one, and two nodes of the new network have an edge connecting them if the corresponding two edges of the original network share a node. In essence, we 'invert' the nodes and edges. It is evident that the two networks are equivalent and either can be retrieved from the other (See Whitney's Line Graph Theorem, \cite{WhitneyTheorem}). Also evident, but perhaps slightly less so, is that the node centrality of the original network is the same as the corresponding edge centrality of the line graph (and vice-versa).
\par 

The metric we propose in this paper is the edge Random Walk Centrality (RWC) \cite{RWCsource}. It "quantifies
how central a node is, regarding its potential to receive information randomly diffusing over the network". In terms of the line graph, this translates to the fact that an edge with higher RWC is likely to receive information before another. The choice of metric for community structure detection may be justified by highlighting the impact of communities on information spread. Tightly-knit communities will have information spread to them faster than loosely-connected nodes.  Below, we reproduce the calculations as given in \cite{RWCsource} for the value of the RWC. Note that the RWC calculated is for the nodes of the given network. However, we shall apply this to the line graphs derived from the networks under our consideration, thus transforming the metric into the edge RWC.
\par

\subsection*{Random Walk Centrality}
We consider the adjacency matrix $\mathbf{A}$ of a finite, undirected network. We define $K$ as the degree distribution vector, i.e., $K_i = \Sigma_j \mathbf{A}_{ij}$ and the total degree of the graph, $N = \Sigma_i K_i$. As in equation 1 of reference \cite{RWCsource}, we have the probability of a random walker starting from node $i$ to be at node $j$ after a time $t$ to be:
\begin{equation}
P_{ij,t+1} = \sum_k \frac{A_{kj}}{K_k}P_{ik,t}
\label{Pmast}
\end{equation}
However, this is subject to the initial condition $P_{ij,0} = \delta_{ij}$, representing the fact that at $t=0$, the signals are all at their starting positions. By inspection, \eqref{Pmast} also gives 
\begin{equation}
P_{ij,t}K_i = P_{ji,t}K_j
\label{Psymm}
\end{equation}
And hence, the infinite time limit (corresponding to the stationary probability distribution $P_i^\infty$ is simply equal to 
\begin{equation}
P_i^\infty = \frac{K_i}{N}
\label{Pinf}
\end{equation}
The characteristic relaxation time $\tau_i$ of the node $i$ is given by the expression
\begin{equation}
\tau_i = \sum_{t=0}^{\infty}(P_{ii,t} - P_i^\infty)
\label{tau}
\end{equation}
And finally, we can write the RWC $C_i$ of node i as 
\begin{equation}
C_i = \frac{P_i^\infty}{\tau_i}
\label{RWC}
\end{equation}
We thus have a centrality metric for the nodes of a given graph. When applied to the line graph of the graph under consideration, it will yield the edge RWC. This is what we shall use in our agglomerative clustering algorithm.
\par

\begin{figure}[!t]
  \includegraphics[scale = 0.85, angle=90,origin=c]{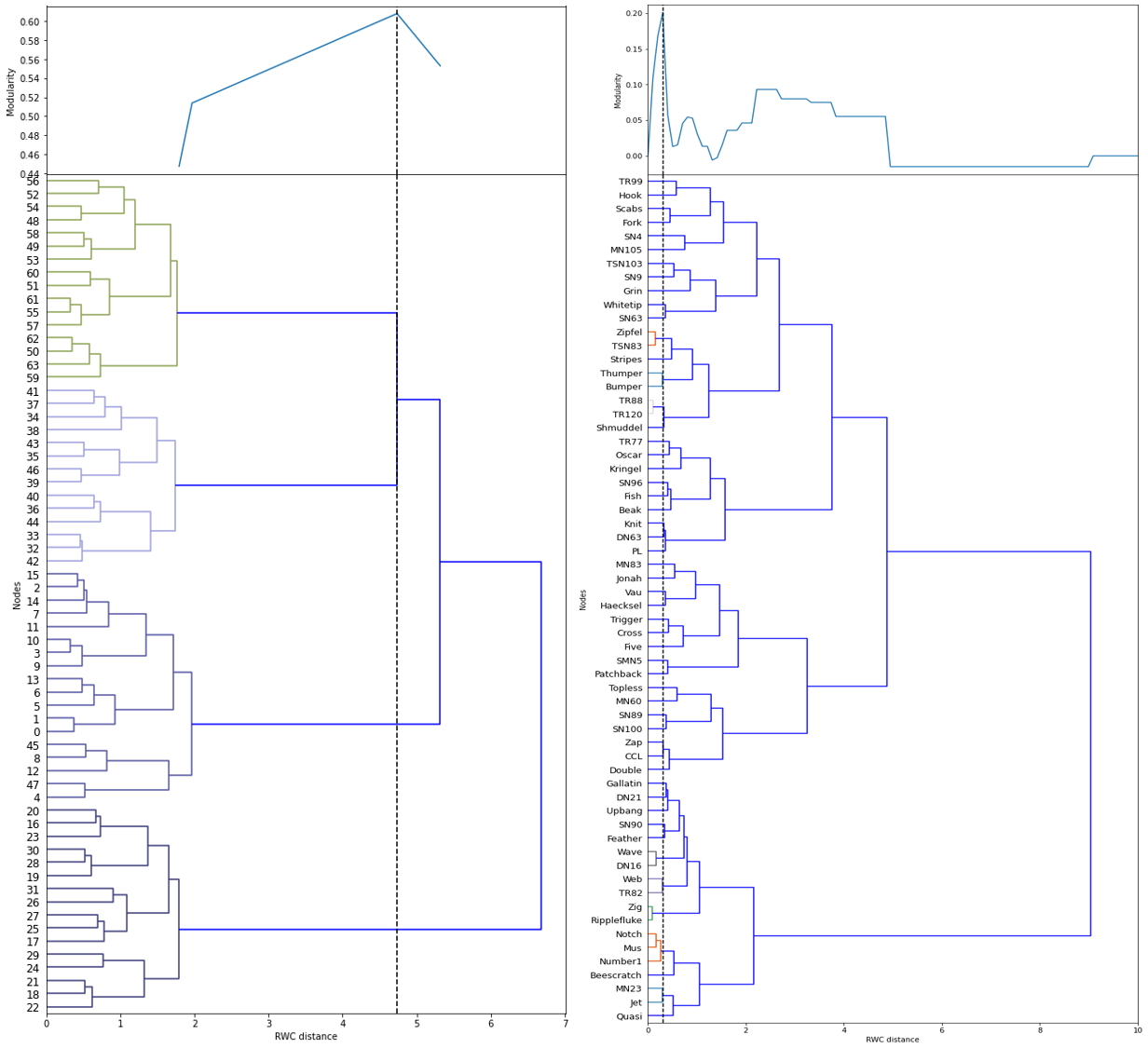}
  \caption{(a) The dendrogram obtained by using Ward's linkage on an artificial network consisting of 4 communities of 16 nodes each. The largest links are those which connect distinct communities. (b) The dendrogram for Lusseau's Dolphin network, clearly showing a pathological peak in modularity very low in the dendrogram - with only 9 links having been added.}
  \label{ArtiDendpathPeak}
\end{figure}

Using the RWC on the line graph for community detection differs from classical random-walk based algorithms in two respects. First, most works define their metrics with respect to nodes \cite{FortunatoReview}, while it is more favorable to use edges. Using edges to identify community structure is motivated by the fact that a 'community' is defined much better in terms of connections (i.e., edges) rather than members (nodes). While a rigorous definition of 'community' is lacking, most modern works agree that a division into communities is good if the "proportion of edges inside the communities is high compared to the proportion of edges between them"\cite{ViaRandomWalk2}. To the same end, the measure for evaluating the quality of community division is the \emph{modularity}, described and first introduced in \cite{GirvanNewman2004} \par
Consider a division of the network into $p$ communities, and define a square, symmetric matrix \textbf{e} of the same size. $e_{ab}$ will denote the fraction of edges that run from community $a$ to community $b$. The modularity is then given by 
\begin{equation}
Q = \Tr\textbf{e} - ||\textbf{e}^2||
\label{modularity}
\end{equation}
where $||\textbf{M}||$ gives the sum of all elements of the matrix $\textbf{M}$. This is exactly what we require: the trace gives the fraction of intra-community edges, while the element sum gives the inter-community edges. \par
 
Second, although line graphs have previously been used to find overlapping community structures (see \cite{OverlappingCommunitiesReview}), using them in this applicative scenario provides a unique advantage - it allows us to investigate directly the edges of the graph using methods well-tested for nodes. The RWC "quantifies how central a node is located regarding its potential to receive information randomly diffusing over the network"\cite{RWCsource}. Applying this on the line graph gives us a measure of how central an \emph{edge} is with respect to information travelling across the network. \par

This paper is organized as follows: In Section II, we expand on the implementation of our algorithm and discuss other computational aspects. In Section III, we present the results of our algorithm when applied to some standard networks and compare them to previous results.  Finally, in Section IV, we showcase an application of our algorithm to the Indian Railway network. \par

\begin{table*}

\begin{tabular}{|c|c|c|c|c|}
    \hline \hline
        Network | Ideal communities & Algorithm & Communities & Correctly classified nodes & Max modularity \\\hline\hline
       \multirow{2}{*}{Football | 12} & RWC & 11 & 99.14\% & 0.13 \\\cline{2-5} 
                                                      & GN & 13 * & 97.41\% & ? \\\cline{1-5}
        \multirow{2}{*}{Karate | 2} & RWC & 2 & 91.18\% & 0.35 \\\cline{2-5}
        & GN & 5 & 70.59\% ** & 0.4 \\\cline{1-5}
        \multirow{2}{*}{Les Miserables | ?} & RWC & 4 & - & 0.4 \\\cline{2-5}
        & GN & 11 & - & 0.54 \\\cline{1-5}
        \multirow{2}{*}{Dolphin | 2} & RWC & 2 & 90.00\% & 0.1 \\\cline{2-5}
        & GN & 5 & N.A \^ & 0.52 \textasciicircum{} \\\cline{1-5}
        \multirow{2}{*}{Collaboration | ?} & RWC & 10 & - & 0.07 \\\cline{2-5}
        & GN & 13 & - & 0.72 \textasciicircum{}\textasciicircum{} \\ \hline \hline
    \end{tabular}
\caption{Comparison of the RWC algorithm and the Girvan Newman algorithm with the ideal (expected) community structure of several real-world networks. The ideal split for \emph{Les Misérables} and Collaboration networks is unknown. The max modularity is undefined for the ideal split.  \newline $^*$Not considering Independent school singleton communities \newline $^{**}$Considering split at the global modularity maximum \newline \textasciicircum{}Girvan and Newman \cite{GirvanNewman2004} used the \cite{DolphinNetwork2} dataset, while we used the \cite{DolphinNetwork} dataset \newline \textasciicircum{}\textasciicircum{}Different datasets were used}
\label{Comparison}
\end{table*}

\section{Implementation}
We now present our hierarchical (agglomerative) clustering algorithm. Given a graph $G$ with $n$ vertices and $m$ edges, perform the following. \newline Preparatory steps:
\begin{enumerate}
\itemsep0em
\item Construct the line graph $G_1$ (which will consist of $m$ vertices and $\sum_{i = 1}^n d_i^2 - m$ edges, where $d_i$ is the degree of vertex $i$). 
\item Calculate the RWC of all the nodes of $G_1$, by \eqref{RWC} and assign these values as weights to the corresponding edges of $G$
\item Define an $n \times n$ distance matrix $R$, such that the entry $R_{ij} $($= R_{ji}$) corresponds to the sum of weights of edges that lie in the (weighted) shortest path from node $i$ to node $j$. This distance matrix then represents the pairwise distance between all nodes of $G$
\end{enumerate}

We now have a forest of nodes (singleton clusters). Iterative steps:
\begin{enumerate}
\itemsep0em
\item Merge the two closest clusters, and replace them with a new cluster
\item Recalculate distances to all other clusters
\end{enumerate}

These steps are repeated until there is only one cluster left. This hierarchical clustering now gives us a dendrogram, which can be analyzed to find the optimal community structure. Concluding steps:
\begin{enumerate}
\itemsep0em
\item Plot modularity \eqref{modularity} against the height at which the dendrogram is cut off for a given range (see below)
\item Obtain the community structure by cutting the dendrogram at the point where modularity reaches its maximum
\end{enumerate}

The 'distance' between clusters (as used above) may be defined in several different ways \cite{Linkages}. These are known as linkage methods, and we believe that using Ward's linkage is the most appropriate. This defines the distance between clusters as increase in the (ESS) error sum of squares. The sum of squares is defined in the usual way, the sum of squared distances of all nodes from their cluster mean. In our analysis on artificially generated networks (Fig. \ref{ArtiDendpathPeak}a), we found that Ward's linkage will give a large distance between clusters if merging them leads to a community structure that is much less pronounced. We thus argue that the optimal community structure will be obtained by cutting the dendrogram at some point of the longest link (i.e., by removing edges which have the largest distance by Ward's linkage). Now there may exist many other links which exist between the endpoints of this longest link. To determine which of these gives us the optimal cutoff, we use the modularity. \par

Calculating modularity in this truncated domain has two advantages: One, it significantly reduces computation time as the number of points at which modularity must be calculated is lower. Second, in some networks (for example the American Football network and Dolphin network (Fig. \ref{ArtiDendpathPeak}b)) we obtain a sharp and narrow spike in modularity at the very beginning of clustering. There are a lot of edges with very similar RWC scores, and in this densely populated stratum of the dendrogram the modularity achieves a peak value with most nodes still remaining singleton communities. Calculating modularity only in the truncated region serves to avoid this pathological peak.\par

However, if the maximum of the modularity is obtained at one of the endpoints of the truncated region, we further calculate the modularity for the next few links on that side, until we reach a local maximum of modularity. This ensures that the truncation does not leave out modularity maxima which are in the same neighborhood. \par

The calculation of the RWC runs in time $\mathcal{O}(n^3)$ for a graph of n nodes. However, since we run the algorithm on the line graph, the complete clustering algorithm takes time $\mathcal{O}(m^3)$, where $m$ is the number of edges in the original network.\par

This method of community detection offers a unique advantage over traditional methods - it eliminates the need of recalculation of the centrality metric at every step. Once the RWC has been calculated for the network, and the distance matrix defined, we no longer need to recalculate the RWC for the new network. The results are comparable to (and in certain cases, even better) than conventional community detection algorithms. The order of standard algorithms like Girvan and Newman's shortest path betweenness algorithm \cite{GirvanNewman2004} is $\mathcal{O}(n^3)$, while ours is $\mathcal{O}(m^3)$, which is the same for sparse graphs. In the next section, we demonstrate this by applying our algorithm to a wide range of standard and well-tested networks. \par

\section{Preliminary tests}

We test our method on graphs whose community structure has been established very firmly. This includes one computer-generated class of graphs and five real world networks: the American College Football network \cite{conferences2000}, Zachary's Karate Club network \cite{KarateClub}, the Bottlenose Dolphin network \cite{DolphinNetwork}, the Les Misérables character network \cite{LesMiserables} and the Network Science coauthorship network \cite{Newman2006}. Our results are summarized in Table \ref{Comparison} \par

\begin{figure*}
  \includegraphics[scale = 0.58]{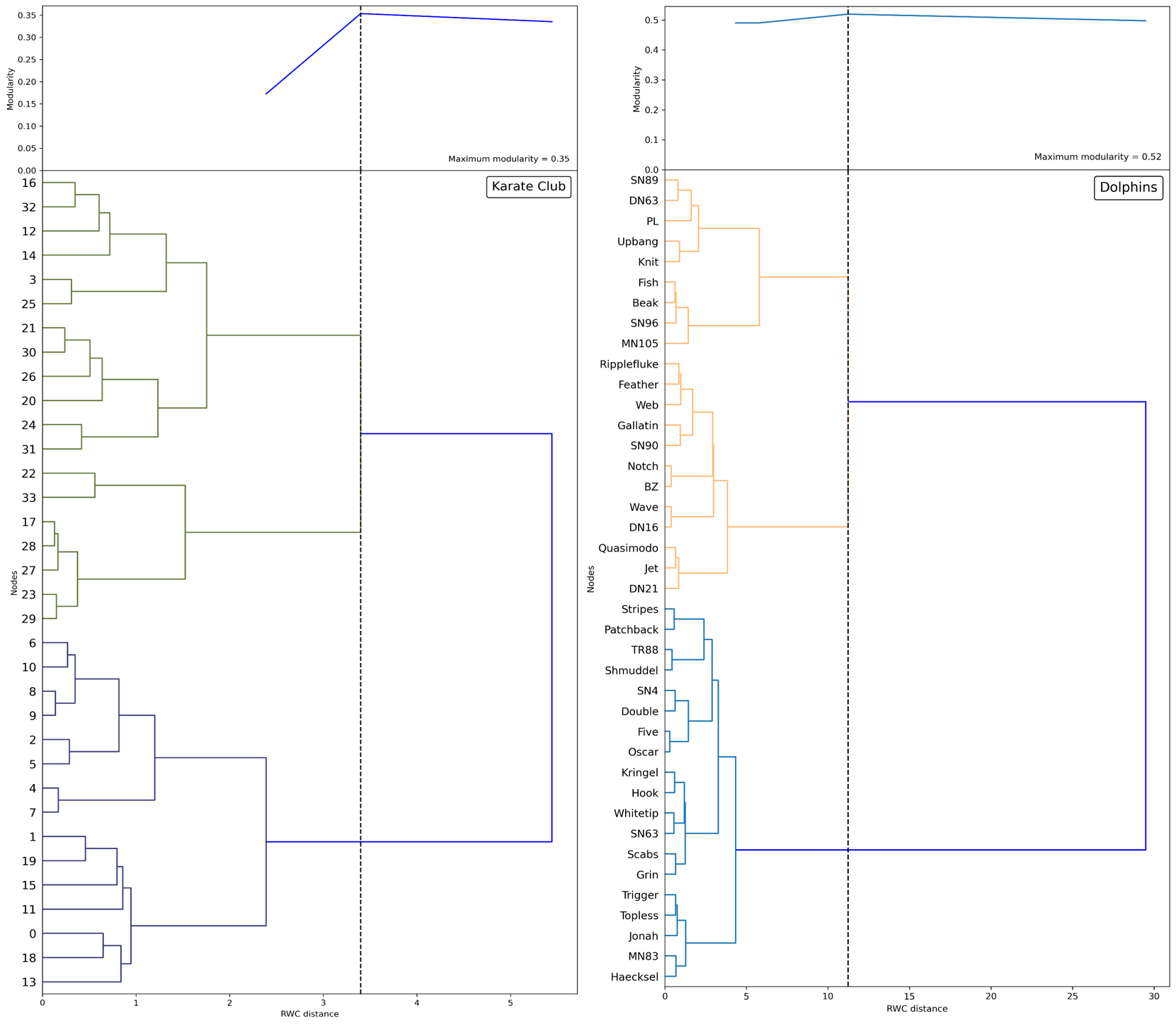}
  \caption{The dendrograms obtained along with the modularity plot in the truncated region, and the corresponding division into communities. (a) shows the split of Zachary's Karate network into 2 communities while (b) shows the split of Lusseau's Dolphin network into 2 groups}
  \label{KarateDolphinsDend}
\end{figure*}

\subsection{Artificial network}

\begin{figure}
\includegraphics{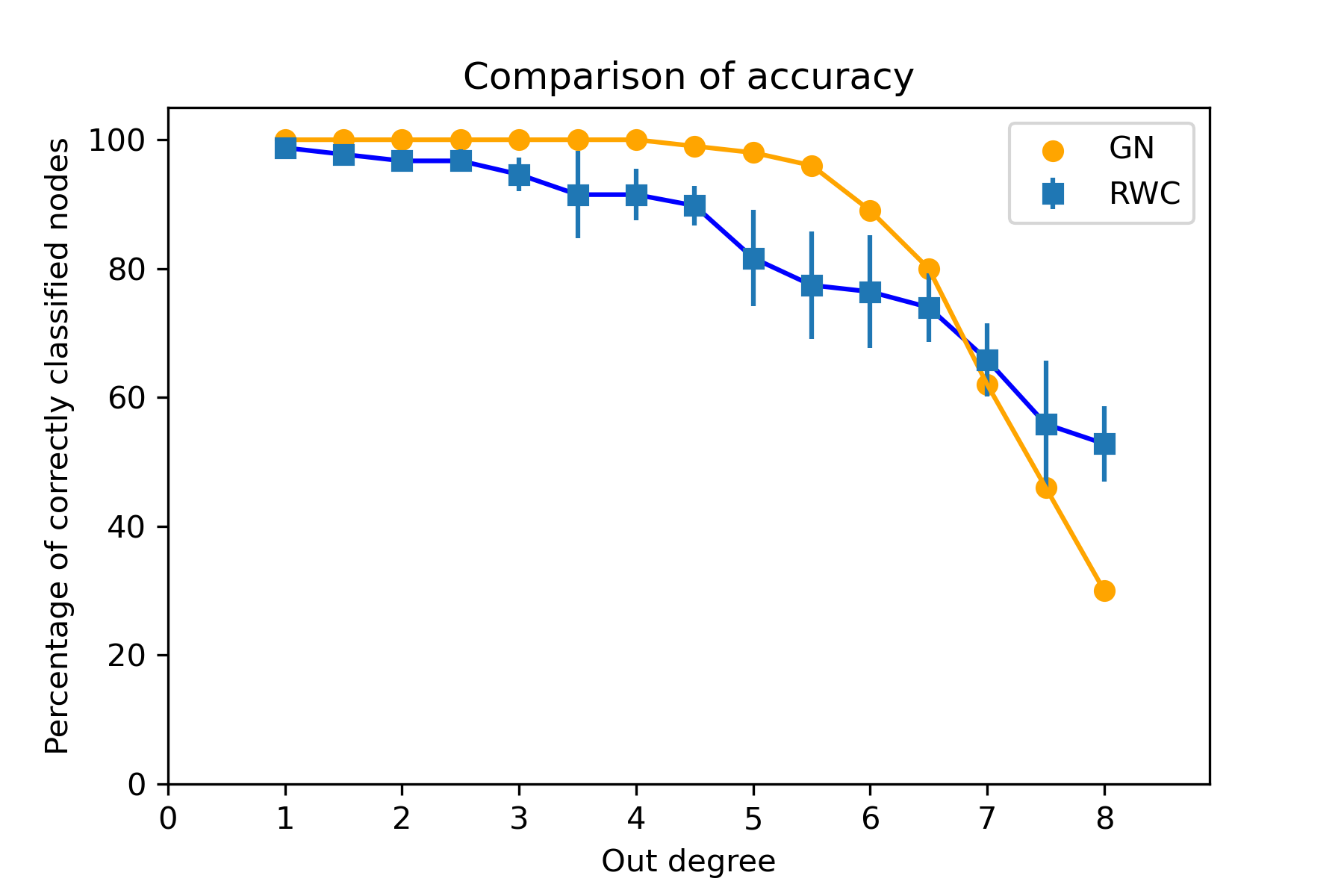}
\caption{The comparison of accuracies of the RWC and Girvan-Newman algorithms. Each point is an average of up to 10 graphs}
\label{Accuracy}
\end{figure}
We created networks with 64 nodes divided into 4 communities of 16 each. The average degree of a node for connection within its community, $z_{\text{in}} = 6$ and we varied the out degree $z_{\text{out}}$ from 0.5 to 5.5, in increments of 0.5. The accuracy of the community division can be found in Fig. \ref{Accuracy}.  While it seems to produce worse results than the Girvan-Newman algorithm for the out-degree range 3 to 7, it appears to perform better at the higher end of the spectrum. This depicts usability in cases where the out degree and in degree are equal and the community structure is very convoluted.
\par

\subsection{American College Football network} 
\begin{figure}
  \includegraphics[scale = 0.54, angle = 90, origin = c]{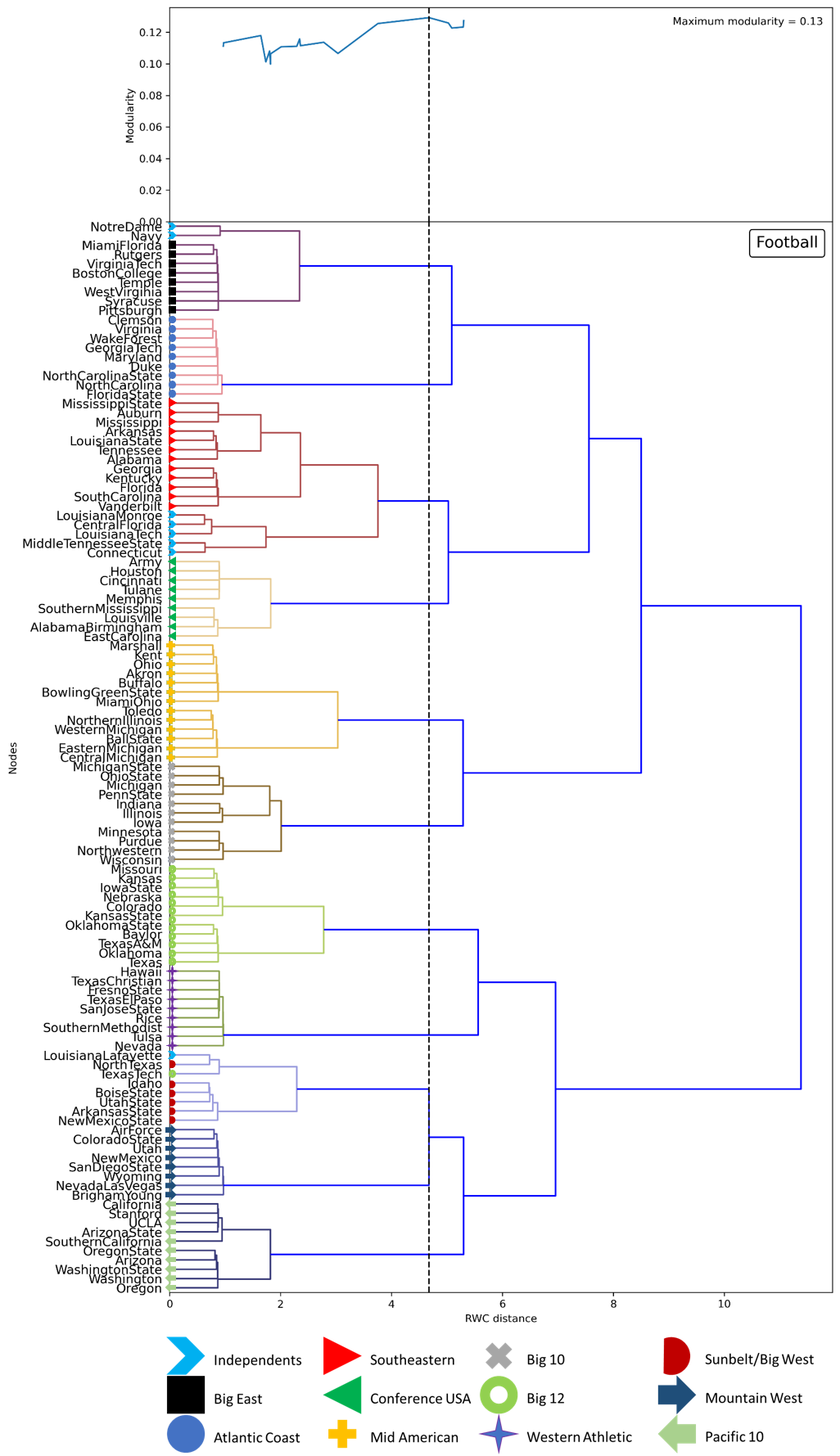}
  \caption[skip=0pt]{The dendrogram for the split of the American 2000 NCAA Division I-A football season. The conference divisions match extremely well with the true conferences, with the exception of Texas Tech. Independents are mostly placed together, except for Notre Dame, Navy and Lousiana Lafayette.}
  \label{FootballDend}
\end{figure}

This network depicts the various American Football college conferences that happened in the USA in 2000. There were a total of 116 schools participating in 12 conferences \cite{conferences2000}. Each node represents a school and each edge represents a game between those two schools. Naturally, the conferences then form communities, and we apply our method to this network. \par
The results are excellent - only one node (Texas Tech) is misclassified. In addition, all the I-A Independent schools occur together in our dendrogram (with the exception of Louisiana Lafayette, Notre Dame and Navy). 
\par

\begin{figure}[ht] 
	\centering 
	\begin{minipage}[t]{6.5cm} 
		\centering 
          \includegraphics[scale = 0.5]{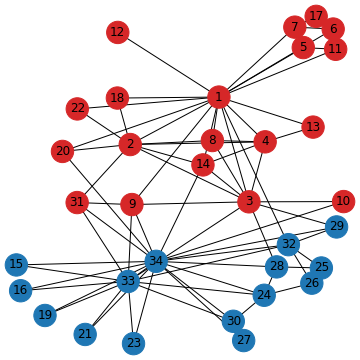}
          \caption{The map showing the split of members of the study of Zachary's Karate club. Node 1 is the administrator of the club, while node 33 is the instructor.}
          \label{KarateMap}
	\end{minipage} 
	\hspace{1cm} 
	\begin{minipage}[t]{6.5cm} 
		\centering 
          \includegraphics[scale = 0.14]{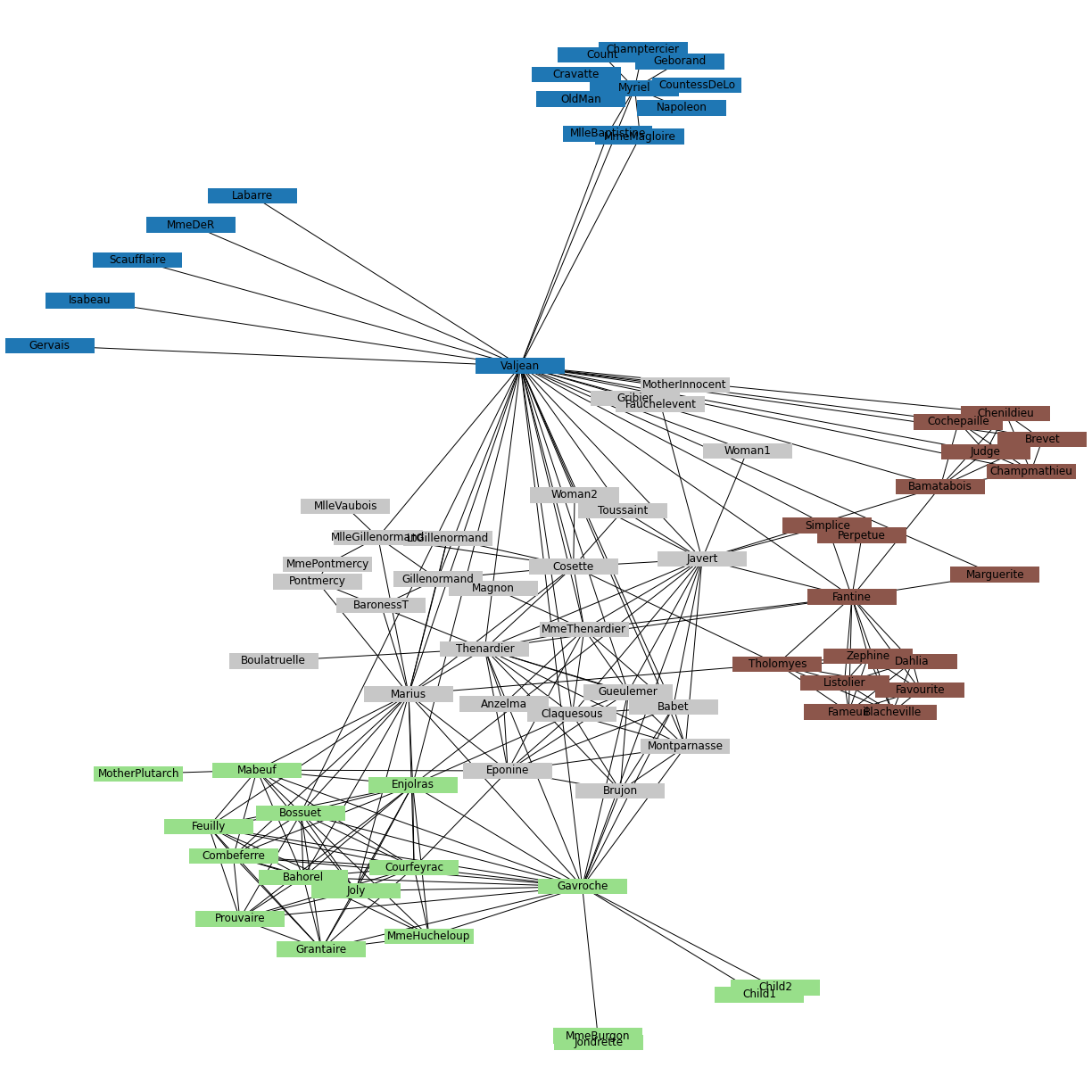}
          \caption{The community map of \emph{Les Misérables}}
          \label{LesmisMap}
	\end{minipage} 
 
\end{figure}

\subsection{Zachary's Karate Club network} 

This is a well-studied network in community detection. Conventional clustering algorithms like the Girvan-Newman usually divide it into two, three or four communities. The 'ideal' community division seems to be that into two groups (stemming from the fact that the club members aligned either with the club's administrator or the instructor after the fission of the club). The RWC algorithm yields a split into 2 groups: very close to the ideal split, and very similar to that of Girvan and Newman \cite{GirvanNewman2004} and the actual split \cite{KarateClub}, with the only difference being the classifications of the nodes at the boundary of the two ideal communities (note that all the three 'misclassified' nodes, 9, 10, and 31 have an equal number of connections to both the ideal communities).
\par

\subsection{Les Misérables character network} 
Victor Hugo's \emph{Les Misérables} presents an ensemble of characters. These characters can be put into a network, with edges representing simultaneous appearance of characters in particular scenes. Different splits have been proposed by various authors (\cite{GirvanNewman2004, localMethods}). However, they all share a common characteristic: Valjean and his adversary Javert form the hubs of two of the largest communities, and the same is observed in the split given by the RWC algorithm (Fig. \ref{LesmisMap}).
\par

\subsection{Bottlenose Dolphin network}
\begin{figure*}
  \includegraphics[scale = 0.45]{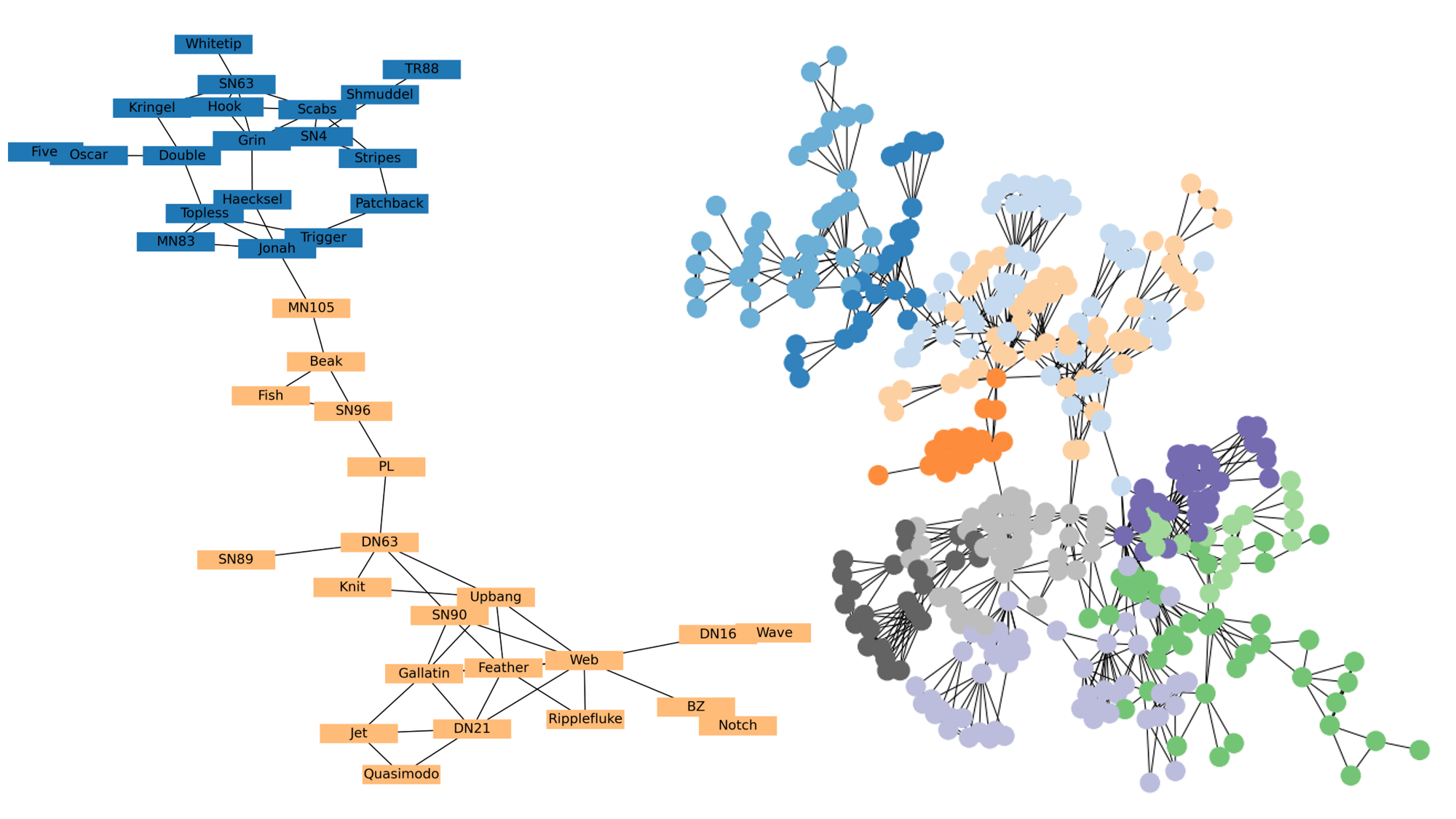}
  \caption{The community map of the Bottlenose Dolphins network (left) and the Network Science Collaboration network (right).}
  \label{DolphinsCoauthorMaps}
\end{figure*}
A study of 62 bottlenose dolphins was conducted by \cite{DolphinNetwork} over seven years. We ran our algorithm on the graph presented in \cite{DolphinNetwork}, and the resulting community structure we found is shown in Fig.  \ref{DolphinsCoauthorMaps}a. In the original paper, the authors identify 3 groups of dolphins. Out of those, one group is claimed to be an artefact (stemming from the low observation frequency of some individuals). This artefact is absent in our network, indicating that it is better than conventional clustering methods. The other two groups match very well to the groups found by our algorithm. The individuals at the boundary of the two groups seem to be placed into the wrong group. However, \cite{GirvanNewman2004} says that the very formation of the two communities occurred because of the temporary disappearance of the individuals at the boundary. In this light, the classification of these individuals doesn't seem as dubious. 
\par

\subsection{Network Science coauthorship network}
This network represents the collaboration between physicists who researched networks, taken from \cite{Newman2006}. The obtained community structure is shown in Fig. \ref{DolphinsCoauthorMaps}. The number of authors is too large for their names to be included, but the division corresponds very well to instiutional affiliations and geographic locations of the authors. 
\par

\section{Indian Railway network}
The Indian Railway Network is one of the most expansive in the world, with about 68,000 km (42000 mi) of track length as of 2022. The graph nodes here have been chosen to correspond with the Divisional Headquarters of the Railways, numbering 70 (Data source: \cite{RailwayData}). The edges correspond to \emph{adjacent} stops that a train makes (i.e., node pairs that are not connected do not imply that there is no train between them, only that there is no direct train between them). We desire that such large networks upon which people rely be well-connected. Below, we list certain characteristics that are expected to be seen in such a well-connected network. \par

\begin{figure*}
  \includegraphics[scale = 0.61]{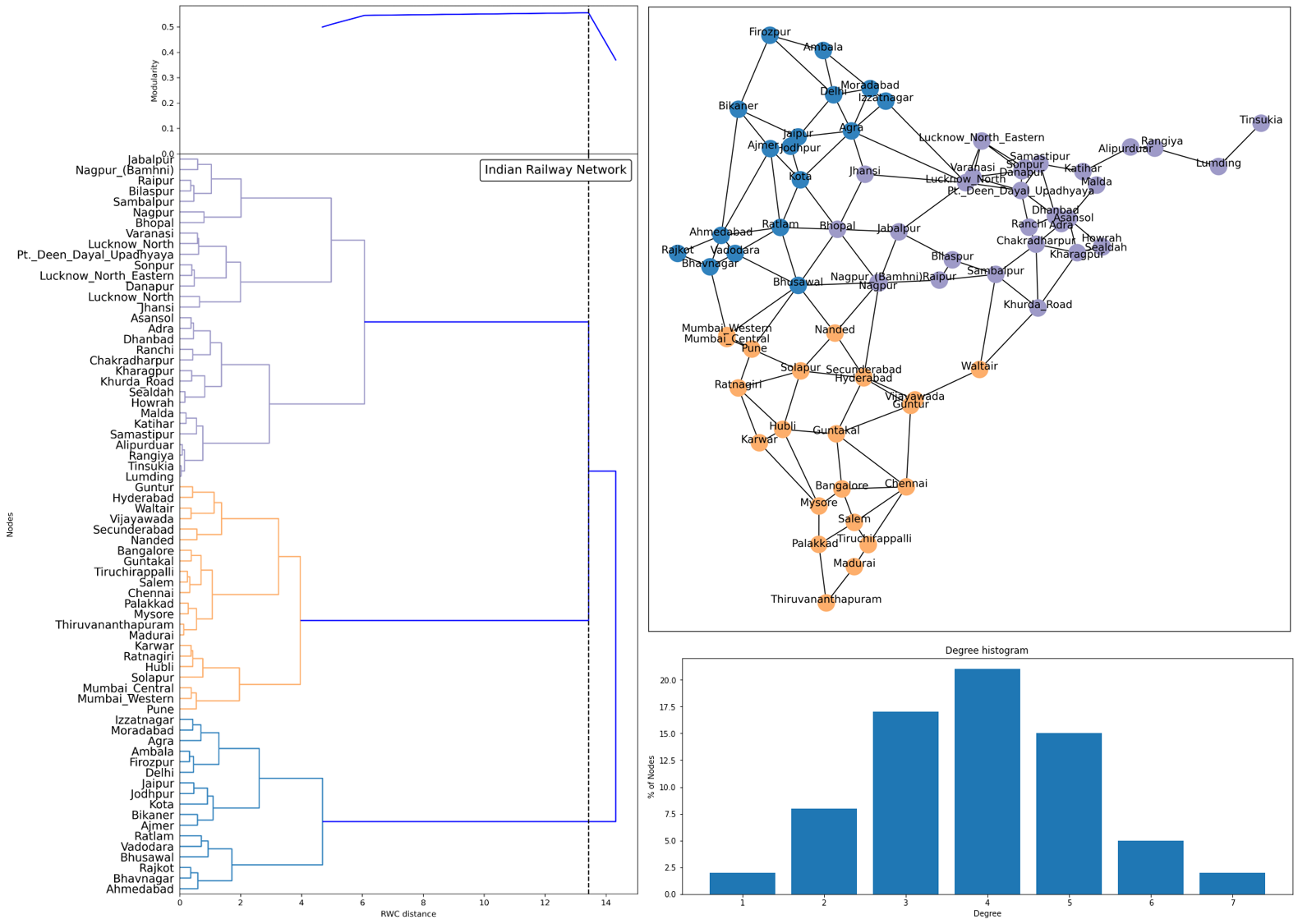}
  \caption{Our results of the analysis of the Indian Railway Network. (a) shows the dendrogram, accompanied by a very flat modularity graph. (b) shows the map of the railway division headquarters, accurate to geographical positions. Three communities are seen, though they are not apparent except for the color assigned to the nodes. (c) shows the distribution of shortest path lengths of the network, clearly showing a single peak.}
  \label{India}
\end{figure*}

\begin{itemize}
\itemsep0em
\item Ideally, a well-connected (sparse) network should have a low number of communities (i.e., the whole graph should be as close to a single community as possible). A high number of communities would mean that nodes lying in different communities are not very well-connected.
\item We do not expect any single link to be very important individually, and this would reflect as a rather flat modularity graph. It would depict the sturdiness of the network to dysfunctional links, which could happen due to a large variety of causes - technical breakdowns, maintenance works, terrorist attacks, traffic overload, etc. A well-connected network must be resilient to these situations to maintain steady flow of traffic and smooth functioning. 
\item A sparse, well-connected network should have a small variance in link importance: all the edges should have approximately equal scores of any given centrality metric. A link which is much too important (or unimportant) would mean a high (low) traffic along it, making it particularly crucial (useless) to the network.
\item It is also reasonable to expect the node density to be fairly uniform over the network. Large fluctuations in concentration of nodes would give rise to community structures, which would not be an expected characteristic of a well-connected network. 
\item Finally, well-connected sparse networks can be expected to show the small-world effect: that the average shortest path between two nodes be much smaller than the size of the network. In popular literature, this shortest path length is found to be near about 6 in most real-world networks, fancily deemed 'six degrees of separation'. If a network is split distinctly into communities, we can expect the shortest path length distribution to show multiple peaks (with the first and largest peak corresponding to node pairs lying in the same community). Higher the number of peaks, the worse is the well-connectedness of the network. 
\end{itemize}

When we apply our community detection algorithm to the Indian Railway Network, these characteristics are exactly what we see. 
\begin{itemize}
\itemsep0em
\item There appear to be only three communities, roughly corresponding to geographically accurate partitions of the country (Fig. \ref{India}b), with the North and West clubbed together, the East as a community and the South as another.
\item The modularity graph (Fig. \ref{India}a) is also nearly constant (in the truncated region), highlighting the resistance of the network to a train that is temporarily not functioning and the presence of alternate routes that are only slightly longer.
\item The lowest stratum of the dendrogram (Fig. \ref{India}a) shows link lengths being very close together (i.e., at the same height).
\item The node density (Fig. \ref{India}b) is not uniform, but given the population density, the node density is found to be in good accordance with it. This reflects the efficient distribution of traffic over nodes, fulfilling the utilitarian well-connected condition.
\item Lastly, the shortest path length distribution of the Indian Railway Network (Fig. \ref{India}c) shows one single peak.
\end{itemize}
Indeed, to make this network even sturdier and more uniform than it already is, we suggest adding more edges (i.e., trains) between nodes lying at the boundary of the three communities (direct trains like Raipur to Waltair/Vijayawada/Guntur/Hyderabad/Secunderabad, Jhansi to Jabalpur/Nagpur/Lucknow, Bhusawal to Karwar/Solapur/Hubli, etc.). Adding these trains also causes the RWC values of the new edges to fall in the same range as the other edges, and no changes in the number of peaks in the shortest path length distribution (with, of course, no change in node density).

\section{Conclusion}
We presented an agglomerative hierarchical algorithm using Noh and Rieger's \cite{RWCsource} Random Walk Centrality as a metric to identify community structure in complex networks. Our approach is unique in three ways. First, the usage of line graphs to find edge centrality gives us the advantage of directly investigating edge connections (which are arguably more fundamental than node linkages in term of community structures). Second, the calculation of RWC only once in the process allows us to run in time $\mathcal{O}(m^3)$ for a graph with $m$ edges. Third, we evaluated the modularity only for the dendrogram stratum defined by the largest edge, when drawn using Ward's Linkage method. This helps reduce computation time without impacting community detection (in fact, it helps surpass pathological peaks in modularity which can sometimes occur when very few edges have been added during clustering). We then tested our algorithm on several standard networks and obtained excellent results. Finally, we demonstrated an application of the algorithm to the Indian Railway Network and checked whether it was well-connected. \par

\section*{Acknowledgements}
The author PM would like to thank the Department of Science and Technology, Government of India, (DST-MATRICS GoI Project No. SERB/F/506/2019-2020 Dated 15th May 2019) for their financial support

\bibliographystyle{plain}
\bibliography{main}

\end{document}